# Influence of strain on magnetization and magnetoelectric effect in La$_{0.7}$A$_{0.3}$MnO$_3$ / PMN-PT(001) (A = Sr; Ca)

C. Thiele,[1] K. Dörr,[1,a)] O. Bilani,[1] J. Rödel,[2] L. Schultz[1]

[1]IFW Dresden, Institute for Metallic Materials, Helmholtzstraße 20, 01069 Dresden, Germany

[2]Institute of Materials Science, University of Technology Dresden, 01062 Dresden, Germany

**Abstract**

We investigate the influence of a well-defined reversible biaxial strain $\leq 0.12$ % on the magnetization ($M$) of epitaxial ferromagnetic manganite films. $M$ has been recorded depending on temperature, strain and magnetic field in 20 - 50 nm thick films. This is accomplished by reversibly compressing the isotropic in-plane lattice parameter of the rhombohedral piezoelectric 0.72PMN-0.28PT (001) substrates by application of an electric field $E \leq 12$ kV cm$^{-1}$. The magnitude of the total variable in-plane strain has been derived. Strain-induced shifts of the ferromagnetic Curie temperature ($T_C$) of up to 19 K were found in La$_{0.7}$Sr$_{0.3}$MnO$_3$ (LSMO) and La$_{0.7}$Ca$_{0.3}$MnO$_3$ films and are quantitatively analysed for LSMO within a cubic model. The observed large magnetoelectric coupling coefficient $\alpha = \mu_0\, dM/dE \leq 6\times10^{-8}$ s m$^{-1}$ at ambient temperature results from the strain-induced $M$ change in the magnetic-film-ferroelectric-substrate system. It corresponds to an enhancement of $\mu_0 \Delta M \leq 19$ mT upon biaxial compression of 0.1 %. The extraordinary large $\alpha$ originates from the combination of three crucial properties: (i) the strong strain dependence of $M$ in the ferromagnetic manganites, (ii) large piezo-strain of the PMN-PT substrates and (iii) effective elastic coupling at the film-substrate interface.

a)Electronic mail: k.doerr@ifw-dresden.de

# I. Introduction

Magnetic ordering in a crystalline solid is sensitive to the bond lengths and angles of the magnetic atoms. This is particularly important for some 3d transition metal oxides characterized by strong electron-phonon interaction. One most-studied example is the family of rare-earth manganites (R,A)MnO$_3$ (R = Y, Bi, La or rare earth metal, A = non-trivalent doping metal) which is well-known for the colossal magnetoresistance (CMR) effect and further extraordinary sensitivities to external parameters including hydrostatic pressure and epitaxial strain in films[1, 2, 3, 4].

Their strong sensitivity towards lattice strain makes manganites interesting candidates for the magnetic part in multiferroic composites with large magnetoelectric effect. Multiferroic composites are understood as a combination of a ferromagnetic and a ferroelectric compound in mixed-powder, layered or nanocolumnar geometries where the components are essentially elastically coupled[5, 6, 7, 8, 9, 10, 11, 12, 13]. The strain induced in one component (either by magnetostriction in the magnet or by inverse piezoelectric effect in the ferroelectric) is mediated to the other and alters its polarization (be it electric or magnetic). Hence, this allows one to control the electric polarization of the composite by a magnetic field or its magnetization ($M$) by an electric field ($E$). The coupling of magnetic and ferroelectric orders in a material is addressed as "magnetoelectric effect" in recent work[14, 15], even though the original term was restricted to single-phase compounds which show a non-zero linear magnetoelectric coefficient $\alpha = \mu_0\, dM/dE$ [16].

In a previous work[12] we have shown that the epitaxial strain in ferromagnetic La$_{0.7}$Sr$_{0.3}$MnO$_3$ films grown on ferro- and piezoelectric Pb(Mg$_{1/3}$Nb$_{2/3}$)$_{0.72}$Ti$_{0.28}$O$_3$ (PMN-PT) (001) crystals can be reversibly tuned by application of an electric voltage to the piezo-crystal. In fact, this approach allows one to control the biaxial strain of epitaxial films as a variable parameter during experiments. Its actual limitation is the achievable magnitude of tunable strain which is ≤ 0.2 % in present experiments. Since PMN-PT(001) is capable of larger strain of at least 0.5 % [17, 18, 19], this limit is not strict but was set for experimental reasons (voltage limit of 500 V across 0.4 mm thick crystals giving $E = 12$ kV cm$^{-1}$, avoidance of crack formation in the piezo-crystal). Note that reversible strain of similar magnitude is applied in mechanical bending experiments[20], but the mechanical apparatus makes bending less versatile than piezoelectric strain control for some experiments. On the other hand, the detailed nature of bending and piezo-strain experiments is different, since bending usually means uniaxial or biaxial expansion associated with a rise in volume. Contrary to this, the present approach is based on piezoelectrically induced biaxial compression of films.

Here, the structural effects on electronic properties in manganites and results for strained thin films are briefly outlined. One well-known structural influence is the reduction of the electronic band width by distortion of the Mn-O-Mn bond angles away from 180°, the bond angle in the cubic perovskite-type lattice. Further, enlarged length of Mn-O bonds also reduces the electronic band width and, as a consequence, the ferromagnetic double exchange interaction[1]. The other important structural effect originates from the strong Jahn-Teller effect of the $Mn^{3+}$ ion in octahedral $MnO_6$ coordination[2]. The degenerate Mn 3d $e_g$ level occupied by one electron is split associated with uniaxial distortion of the surrounding O octahedron. The distortion favours one certain $e_g$ electron orbital and may occur as long-range orbital ordering phenomenon[21, 22, 23]. The epitaxial strain in films was predicted to have a similar effect, since it enforces long-range (tetragonal) distortion. Indeed, strong impact of strain on orbital ordering in thin films has been observed experimentally (e. g. Refs. 24, 25). In ferromagnetic conducting manganites like $La_{0.7}A_{0.3}MnO_3$ (A = Sr; Ca; Ba), the Jahn-Teller distortions are dynamic due to moving $e_g$ electrons[2, 26, 27], but epitaxial strain favours one direction of distortions. Numerous investigations have been carried out studying ferromagnetic manganite films of various thickness deposited on mismatching substrates (e. g. Refs. 28, 29, 30, 31, 32). Typically, the elastic strain in films decreases with increasing film thickness, the decrease being either continuous or abrupt depending on the strain-relaxing defects formed in the film. The ferromagnetic Curie temperature $T_C$ of $La_{0.7}Sr_{0.3}MnO_3$ (LSMO) is reduced by epitaxial strain by up to 100 K [28, 29], whereas $La_{0.7}Ca_{0.3}MnO_3$ (LCMO) shows an enhanced $T_C$ in a certain range of compressive strain. The reason for the difference is attributed to the nearly cubic lattice of LSMO in contrast to the orthorhombic lattice of bulk LCMO. Further, a strong impact of film strain on magnetic anisotropy has been revealed in numerous studies (e. g. Refs. 30, 32, 33, 34, 35). Stress-induced anisotropy is typically dominating over the weak magnetocrystalline anisotropy in nearly cubic lanthanum manganites, with the well-known example of perpendicular magnetization in compressively strained $La_{0.7}Sr_{0.3}MnO_3$ films[36]. However, it is important to note that ferromagnetic ordering is also affected by the finite thickness of films and the oxygen content. The latter may depend on the strain state of the film and is impossible to measure with the needed accuracy thus far. Hence, a dynamic strain experiment is useful in order to clarify the sole influence of strain on magnetic ordering.

In this work, the influence of a well-defined piezoelectrically controlled biaxial compression on the magnetization of epitaxial $La_{0.7}Sr_{0.3}MnO_3$ and $La_{0.7}Ca_{0.3}MnO_3$ films is reported. Substantial strain-induced magnetization is observed also in the remanent state. Magnetoelectric coupling coefficients $\alpha$ of up to $6 \times 10^{-8}$ s m$^{-1}$ at ambient temperature have been revealed. Sections A, B and C of the Results paragraph describe the measured magnetization depending on temperature, biaxial strain and magnetic field at varied strain states. The origin of the strain influence on magnetization is addressed in Sec.D. Sec. E discusses the quantitative determination of the tunable film strain. Finally, the biaxial strain dependence of $T_C$ is analysed for a $La_{0.7}Sr_{0.3}MnO_3$ film following the model proposed by Millis et al.[37] in Sec.F.

**II. Experiment**

Thin epitaxial films of La$_{0.7}$A$_{0.3}$MnO$_3$ (A = Sr; Ca) have been grown by pulsed laser deposition (PLD, KrF 248 nm excimer laser) at 650°C on monocrystalline platelets of rhombohedral Pb(Mg$_{1/3}$Nb$_{2/3}$)$_{0.72}$Ti$_{0.28}$O$_3$(001) (PMN-PT) as described previously[12]. The in-plane lattice parameter of the PMN-PT crystals is $a = b =$ 4.022 Å [38], leading to tensile strain in the films which have pseudocubic bulk lattice parameters of $a_{LSMO} =$ 3.876 Å and $a_{LCMO} =$ 3.864 Å. The film thickness is between 20 nm (50 pseudocubic unit cells) and 50 nm for this investigation. This range has been chosen in order to probe bulk-like behaviour (> 20 nm), but avoid the stronger strain inhomogeneity expected in thicker films.

As-grown films are characterized by $\theta - 2\theta$ x-ray diffraction, atomic force microscopy (AFM) and measurements of electrical resistance. The magnetization ($M$) has been measured in a SQUID magnetometer (Quantum Design) along the [100] direction of the substrate. Either a small magnetic field is applied in this direction during the measurement or the remanent magnetization is recorded after initial application of 5 T. In the latter case, the negligibility of relaxation has been checked.

The epitaxial strain in the films is controlled as follows[12]: The piezoelectric PMN-PT platelets have a NiCr/Au electrode on one (001) face, and the conducting manganite film on the opposite (001) face serves as second electrode. Even though a manganite film may have a thousand-fold higher resistance than the Au electrode, the huge resistance of the PMN-PT platelet of > 1 G$\Omega$ guarantees proper function of the oxide electrode. Between the electrodes, an electric voltage up to 500 V is applied producing an electric field $E \leq$ 12 kV cm$^{-1}$ in the 0.4 mm thick piezo-crystal. Fig.1 shows the reversible in-plane strain of a representative substrate crystal recorded by laser interferometry. Upon application of $E$, both in-plane directions of the substrate shrink approximately linearly with increasing $E$. Despite the presence of ferroelectric domains, the strain is sufficiently uniform in the (001) film plane (cf. Sec. E). In order to avoid both hysteresis and risk of cracks due to mechanical forces at polarization reversal, most experiments are carried out at $E_{max} > E > 0$ after applying $E_{max}$ to the crystal. The temperature dependence of the piezoelectric strain is assumed to be sufficiently weak[39] to be neglected here, apart from an increase of the ferroelectric coercive fields at low temperature. Further, the negligibility of the magnetic signal from the current flowing through the piezo-crystal (limited to 1 $\mu$A, but usually being < 0.1 $\mu$A) has been verified in the paramagnetic state of the films.

**III. Results and Discussion**

**A. Temperature-dependent magnetization at controlled strain**

Tab.1 summarizes the sample number, compound, thickness, out-of-plane lattice parameter and $T_C$ of the investigated films in as-grown state. The temperature dependence $M(T)$ of a La$_{0.7}$Sr$_{0.3}$MnO$_3$ film (#2) and a La$_{0.7}$Ca$_{0.3}$MnO$_3$ film (#4) in remanent magnetic state at $E = 0$ are shown in Figs.2a and 2c, respectively. Both films are under tensile strain (cf. $c$ parameters, Tab.1) from the substrate, leading in combination with finite thickness to lowered values of $T_C$. The bulk $T_C$ of LSMO (LCMO) is ~370 K (250 K) [40, 3]. For both films, the quadratic magnetization $M^2(T)$ (Figs.2b,d) is approximately linear in a temperature range below $T_C$. Hence, the value of $T_C$ in mean-field approximation has been derived by extrapolating the linear part of $M^2(T)$ to $M = 0$. Close to and above $T_C$, $M(T)$ is characterized by a weaker drop (a "tail") towards higher temperature, reminding one of the short range order phenomenon discussed for manganites[3]. Since the "tail" is also observed for remanent magnetization, it is not induced by the magnetic field.

The tensile strain of the as-grown films is reduced by application of $E$ to the substrate. $M(T)$ curves recorded at constant piezoelectric strain are displayed in Figs.2b,d. $M$ increases upon film compression due to the release of tensile strain. The shift of $T_C$ with the strain has been derived; Tab.1 lists the increase of $T_C$ obtained after piezoelectric compression ($\delta\varepsilon_{xx}$) of three films. The largest shift of $T_C$ of 19 K is recorded for the thinnest (20 nm) LSMO film (#2) which has a much suppressed $T_C$ (278 K). The quantitative relation of $T_C$ vs. strain derived in Sec. F shows that the reduced $T_C$ of this film cannot be explained by its strain state alone. Some additional reduction of $T_C$ is ascribed to its finite thickness and / or strain-induced under-oxygenation. The La$_{0.7}$Ca$_{0.3}$MnO$_3$ film also has a strongly suppressed $T_C$, but its absolute strain-induced shift of $\Delta T_C$ = 11 K is not higher than that of the LSMO films.

Most efficient strain control of $M$ is possible close to $T_C$ if the strain effect is essentially based on changing $T_C$ (cf. Sec.D). (One further approach for strain control of $M$ is the change of magnetic anisotropy, but this plays a minor role in the present experiment, cf. Sec. C.) For applications, one may think of creating a distribution of $T_C$ values in a film in order to achieve large strain-dependent $M$ in a wide temperature range. On the other hand, even strain-induced paramagnetic-ferromagnetic switching at well-chosen temperatures should be possible in a film with "sharp" $T_C$. For the present films, this is hampered by the "tail" of $M(T)$ towards $T > T_C$.

**B. Strain control of magnetization**

The magnetization of a La$_{0.7}$Sr$_{0.3}$MnO$_3$ film (#3, Fig.3a) is enhanced by about 25 % at 330 K when $E = 10$ kV cm$^{-1}$ has been applied. The $M(E)$ loop of the LCMO film (#4, Fig.3c) confirms that the remanent $M$ can also be controlled in this way. Interestingly, the strain-dependent $M$ loop is hysteretic up to maximum $E$, in contrast to the non-hysteretic strain of the substrate above the ferroelectric coercive field $E_C$ ~ 2 kV cm$^{-1}$ (Fig.1). This may indicate a magnetic origin of the hysteresis which is not unlikely since the film is in a multi-domain state at the measuring field. One origin of strain-induced $M$ has been revealed in Sec. A, i. e. the enhancement of $T_C$ by the piezoelectric compression. Change of magnetic anisotropy as another possible source is discussed in Sec. C.

The direct strain control of $M$ may be of practical interest for electrically tuning a permanent magnet. Fig.3b illustrates the derived magnetoelectric coupling coefficient $\alpha = \mu_0 \, dM/dE$ vs. electric field for the film of Fig.3a. The value of $\alpha \leq 6 \times 10^{-8}$ s m$^{-1}$ is larger than those values observed for single-phase materials and earlier-investigated composites[14]. It could be comparable to the response expected in layered composites of PMN-PT and magnetostrictive materials[41], but for these samples which show the largest known magnetoelectric voltage coefficient $\alpha^* = dE/dH$ the change of $M$ is not yet investigated. The large value of $\alpha$ indicates effective elastic coupling at the epitaxial film-substrate interface. Assuming ideal elastic interface coupling, $\alpha$ can be estimated as $\alpha = \mu_0 dM/d\varepsilon_{xx} \cdot d\varepsilon_{xx}/dE$ with the in-plane strain $\varepsilon_{xx}$, the inverse piezoelectric effect $d\varepsilon_{xx}/dE$ of the substrate and the strain coefficient $\mu_0 \, dM/d\varepsilon_{xx}$ of the magnetic film. Both factors $dM/d\varepsilon_{xx}$ and $d\varepsilon_{xx}/dE$ can be optimized independently by choosing a strain-sensitive magnet and a ferroelectric with large inverse piezoelectric response.

**C. Magnetization vs. magnetic field at controlled strain**

Field-dependent demagnetization curves of a LSMO film (#3) have been studied close to $T_C$ and at 10 K (Fig.4). The easy axes are the in-plane diagonals [110] and [1$\underline{1}$0], as visible in Fig.4a. This is typical for La$_{0.7}$Sr$_{0.3}$MnO$_3$ films under isotropic tensile stress, e. g. for films on SrTiO$_3$(001) [4, 33]. Hence, most measurements presented here have been taken along a [100] hard axis. At low temperature, the observed influence of the piezo-strain is low: no detectable strain-dependent change of the magnetic coercive field $H_{Cm}$ and negligible enhancement of the saturated magnetization and saturation field (Fig.4b). This is, nevertheless, qualitatively in line with experiments on statically strained films, because the strain does not change the angle between the easy axes and the measuring direction in our experiment, and the magnitude of $\delta\varepsilon_{xx} \sim -0.1\,\%$ is probably too small to substantially modify $H_{Cm}$ or the saturation field of $M$. Hence, strain-induced modulation of magnetic anisotropy is concluded to play a minor role for the reported experiment.

**D. Origin of the strain influence on magnetization**

As visible in Fig.4c, the saturated magnetization at ambient temperature depends on strain. The saturated magnetization equals the spontaneous magnetization $M_S$ in a ferromagnet. Here, we address the strain influence on $M_S$ whereas possible strain effects on domain processes are beyond the scope of this work. (They are not expected to dominate the observed behaviour at present strain levels.) Kuzmin et al. [42, 43] reported on a universal function which describes the reduced spontaneous magnetization $m = M_S/M_S(T=0)$ of ferromagnets in the full temperature range $0 \leq \tau = T/T_C \leq 1$. This expression is considered for a phenomenological discussion of strain-dependent $M_S$:

$$m(\tau) = \left[ 1 - s\tau^{3/2} - (1-s)\tau^{5/2} \right]^{\beta} \tag{1}$$

with $\beta \approx 1/3$ and $s$ as a free (shape) parameter. The three parameters $T_C$, $M_S(T=0)$ and $s$ can principally be affected by strain. Figs.4a,b have shown that $M_S(T=0)$ remains rather unchanged by strain. Note that this is expected if spin-only magnetic moments of Mn ions align in parallel. Unfortunately, it is not yet possible to conclude about the parameter $s$, because the temperature range of recorded $M(T)$ data was restricted to the mean-field-like range below $T_C$. The strain-dependent shift of $T_C$ is clearly detected (Sec. A), hence it is considered as essential origin of the strain-dependent $M_S$.

At this point, the question about the microscopic origin of the strain-dependent $T_C$ arises. Direct calculations of $T_C$ vs. a tetragonal distortion for $La_{0.7}A_{0.3}MnO_3$ are not yet known to the authors. A number of studies have been devoted to half-doped insulating manganites and the issue of strain stabilization of orbital ordering[21, 22, 23]. Millis et al.[2, 37] have revealed the crucial balance between the total energy gain from local Jahn-Teller distortion induced by charge carriers and the kinetic energy of electrons for the conduction mechanism coupled with ferromagnetism. Fang et al. derived the stable magnetic phases in the plane of tetragonal distortion $c/a$ vs. doping ($x$) for La$_{1-x}$Sr$_x$MnO$_3$ from first-principles band structure calculations. They obtained a stable ferromagnetic phase at $x = 0.3$ for $c/a = 1$ which changes to layered antiferromagnetic for strong tensile strain ($c/a < 0.96$). This magnitude of strain is not reached in the present experiment with $c/a \geq 0.975$ in as-grown LSMO films, in consistence with their ferromagnetic behaviour. The computational result just qualitatively confirms the weakening of ferromagnetism with increasing tensile strain which stabilizes the in-plane $e_g$ electron orbital.

The local structural response towards epitaxial strain in a film may include: (i) a change in the average Mn-O bond length, (ii) a change in the Mn-O-Mn bond angles and (iii) a distortion of the O octahedra influencing directly the $e_g$ level splitting. (i) and (ii) affect the electron hopping integral and, hence, the kinetic energy. Experimentally, there is a growing number of studies of the local distortions induced by film strain. Booth et al.[26, 44] employed x-ray absorption techniques (XAFS); their studies of the variance of Mn-O bond lengths demonstrated the dynamic distortions in metallic manganites (in their work, La$_{1-x}$Ca$_x$MnO$_3$). Further investigations using x-ray absorption techniques have been conducted on strained films of La$_{1-x}$Ca$_x$MnO$_3$ [45], Nd$_{0.5}$Sr$_{0.5}$MnO$_3$ [46] and La$_{0.7}$Sr$_{0.3}$MnO$_3$ [47]. They differ in their conclusions regarding the relevance of changes in bond angles, average bond length and shape of O octahedra. This may indicate a specific reaction towards strain depending on the composition / bulk lattice structure of the respective manganite. Optical reflectance / transmittance spectra give evidence for strain-induced shifts of phonon modes[31]. The available results for La$_{0.7}$Sr$_{0.3}$MnO$_3$ films[31, 47] indicate a distortion of oxygen octahedra as a major structural response to tetragonal strain. However, more investigations are needed for a clear picture.

**E. Estimation of strain-controlled lattice parameters**

First, the uniformity of the in-plane lattice parameter of 0.72PMN-0.28PT(001) as a crucial issue for quantitative strain analysis is considered. Ferroelectrics are generally non-cubic. They form ferroelectric domains which typically lead to varying lattice parameters on a crystal surface. Advantageously, rhombohedral (or tetragonal) crystals polarised along [001] have a constant value of the in-plane lattice parameter on (001) surfaces. For rhombohedral crystals, the local polarization oriented along one of the four pseudocubic space diagonals [111], [$\underline{1}$11] etc. has the same tilt angle with the surface everywhere. Deviations only arise at the narrow ferroelectric domain walls where the polarization direction changes. However, the domain walls in comparable oxide ferroelectrics are only few unit cells wide, i. e. they affect only a small volume fraction of a film grown on top. Above the domain walls, small kinks appear at the PMN-PT(001) surface as imaged by AFM line scans[48]. The angle of the kinks is between 179° and 180° for a large number of investigated domain walls. It is related to the small rhombohedral distortion of 0.72PMN-0.28PT, with the rhombohedral angle of 89.90° at 300 K obtained from refined x-ray data on a powderized crystal[38]. It is useful to note that *the rhombohedral distortion of PMN-PT (x = 0.28) is comparable to that of LaAlO$_3$* [49] which is frequently used for epitaxial growth of oxide films. Hence, we suggest that the deviations from cubic lattice structure are sufficiently small for 0.72PMN-0.28PT to justify the assumption of a uniform in-plane lattice parameter for most experiments. Upon application of an electric field (which is below the critical field inducing a rhombohedral-tetragonal transition in PMN-PT [17]), a tetragonal distortion takes place which maintains the uniformity of the in-plane lattice parameter.

The piezoelectric substrate strain is assumed to be fully transferred to the thin film. Hence, the film undergoes a biaxial compression of known magnitude. The film deformation in the case of manganites has been found to be elastic up to the applied compression of -0.25 %, since no irreversible changes of resistance were detected. The lower strain compared to the nominal film-substrate mismatch (3.1 %) of as-grown films (Tab.1, $\varepsilon_{zz}$) reveals that our films contain lattice defects. It is difficult to estimate the influence of these defects on the transfer of piezoelectric strain perpendicular to the film normal; for this reason, this investigation has been restricted to a film thickness ≤ 50 nm. The total strain of La$_{0.7}$Sr$_{0.3}$MnO$_3$ films in dependence on $E$ is derived from the film`s $c$ lattice parameter in as-grown state, the piezoelectric substrate in-plane strain ($\delta\varepsilon_{xx}$) and the Poisson number $\nu \sim 0.33$ (which is known for this compound from several studies[28, 29, 30] and, hence, considered to be sufficiently reliable) as

$$\varepsilon_{xx}(E) = -\varepsilon_{zz}(E=0) + \delta\varepsilon_{xx}(E) \quad \text{(with } \delta\varepsilon_{xx} < 0 \text{ and } \varepsilon_{xx} = -\varepsilon_{zz} \text{ for } \nu = 0.33\text{)} \quad (2)$$

In future work, the film in-plane and out-of-plane lattice parameters might be determined from x-ray diffraction with electric voltage applied to the substrate. In this way, also the Poisson number of thin films might be directly measurable. For LCMO, no reliable Poisson number seems to be available, and it may be anisotropic.

After estimation of $\varepsilon_{xx}(E)$, $M$ data can be plotted vs. in-plane strain, as shown in Fig.5 for one branch of decreasing electric field from Fig.3a. Interestingly, the dependence $M(\varepsilon_{xx})$ is clearly linear in the investigated range of $\varepsilon_{xx}$ for LSMO even though both the strain vs. $E$ (Fig.1) and $M(E)$ (Fig.3a) show some non-linearity. The linear strain coefficient of $M$ is derived as $dM/d\varepsilon_{xx} = 1.3 \times 10^4$ emu cm$^{-3}$ (with absolute values of strain like $\varepsilon_{xx} = 1.2 \times 10^{-3}$), or in other units $\mu_0 \, dM/d\varepsilon_{xx} = 19$ T (leading to the strain-induced $\mu_0 M = 190$ mT by 1 % of compression).

### F. Quantitative strain dependence of $T_C$ in La$_{0.7}$Sr$_{0.3}$MnO$_3$

In Figs.6a,b, $T_C$ of film #1 is plotted vs. $E$ and vs. the total in-plane strain $\varepsilon_{xx}$. The scattering of the data is essentially caused by the error of $T_C$ estimation (cf. Sec. A). The shift of $T_C$ resulting from an elastic tetragonal deformation of the cubic unit cell can be expressed according to Millis et al.[37] as

$$T_C = T_{C0}\left(1 + \alpha\varepsilon_B - \frac{\Delta}{2}\left(\varepsilon^*\right)^2\right) \quad (3)$$

with the bulk compression $\varepsilon_B = -1/3(\varepsilon_{xx} + \varepsilon_{yy} + \varepsilon_{zz})$ and the isotropic biaxial distortion $\varepsilon^* = 1/2(\varepsilon_{zz} - \varepsilon_{xx})$. $\varepsilon_{xx}$, $\varepsilon_{yy}$ and $\varepsilon_{zz}$ are the diagonal elements of the strain tensor with $\varepsilon_{xx} = \varepsilon_{yy}$ for isotropic in-plane distortion. The shear strains $\varepsilon_{xy}$ etc. are negligible if no bending occurs. This is assumed to hold for our experiment due to the low thickness of both electrode layers on the piezo-crystal. One may derive the bulk coefficient $\alpha = 1/T_C \, dT_C/d\varepsilon_B = -3/T_C \, dT_C/dp \cdot (1/V \, dV/dp)^{-1}$ from the pressure ($p$) dependence of $T_C$ and the volume ($V$) compressibility. The parameter $\Delta = 1/T_C \, d^2T_C/d(\varepsilon^*)^2$ characterizes the sensitivity of magnetic ordering towards biaxial strain. In a cubic lattice, the linear derivative of $T_C$ with respect to $\varepsilon^*$ is zero[37]. Under the assumption of a known Poisson number of $\nu = 0.33$, eq.(3) is reduced to

$$T_C = T_{C0}\left(1 - \frac{\alpha}{3}\varepsilon_{xx} - \frac{\Delta}{2}\varepsilon_{xx}^2\right) \qquad (4)$$

With a good data set covering an extended range of in-plane strain $\varepsilon_{xx}$, all three parameters $T_{C0}$, $\alpha$ and $\Delta$ could in principle be fitted. Since our data cover a rather small range of strain and show some scattering of $T_C$ values, the volume parameter $\alpha$ has been derived from neutron work on a crystal of the same composition[50]. That work records the pressure dependences of the volume at ambient temperature ($5.4 \times 10^{-3}$ GPa$^{-1}$) and of $T_C$ (1.2 % GPa$^{-1}$). The latter value has an error of up to 20 % according to the authors, but agrees fairly well with cited earlier studies. Hence, $\alpha = 6 \pm 1.2$ for La$_{0.7}$Sr$_{0.3}$MnO$_3$. Note that the volume term in eqs.(3, 4) contributes about 10 % of the measured shift of $T_C$ in the present experiment; it is not negligible but its error is less crucial. The data of Fig.6b can be fitted with eq.(4) and yield $T_{C0} \sim 370$ K and $\Delta \sim 2000$. First, this result supports the reliability of the experiment, since $T_{C0}$ is close to the bulk $T_C$. The value of $\Delta$ compares in the order of magnitude with results on statically strained La$_{0.7}$Sr$_{0.3}$MnO$_3$ films (Angeloni et al.[29]: $\Delta = 1000$; Ranno et al.[30] : $\Delta = 2100$; Lu et al. Lu[51]: $\Delta = 1400$ for La$_{0.67}$Ba$_{0.33}$MnO$_3$) which, on the other hand, vary by a factor of two.

For the Ca-doped film, the elastic response of the orthorhombic material is probably not describable with a cubic model, since even the volume compression observed under hydrostatic pressure is anisotropic[52]. Hence, quantitative analysis of the biaxial strain dependence of $T_C$ in La$_{0.7}$Ca$_{0.3}$MnO$_3$ may need to take anisotropy into account and is beyond this work.

**IV. Conclusion**

The influence of biaxial strain on the magnetization of epitaxial films of two prototype ferromagnetic manganites has been quantitatively investigated. This is accomplished by reversibly compressing the isotropic in-plane lattice parameter of rhombohedral 0.72PMN-0.28PT (001) substrates by application of an electric field $E \leq$ 12 kV cm$^{-1}$. Biaxial compression of the epitaxial films induced an enhanced magnetization of the order of $\mu_0 \Delta M / |\delta \varepsilon_{xx}| \leq$ 190 mT % $^{-1}$ in low or zero magnetic field. The inspection of temperature-dependent $M$ data at varied strain level reveals a pronounced strain-induced shift of $T_C$ and low strain influence on zero-temperature saturated magnetization. The biaxial strain parameter[37] $\Delta$ = 2000 is derived from strain-dependent $T_C$ data for La$_{0.7}$Sr$_{0.3}$MnO$_3$. A large strain-mediated magnetoelectric coupling coefficient $\alpha = \mu_0 \, dM/dE \leq 6\times10^{-8}$ s m$^{-1}$ at ambient temperature is observed for the magnetic-film-ferroelectric-substrate system. Hence, electrical control of a permanent magnet at practically relevant temperature has been demonstrated.

In the final state of this work we got aware of the work by Eerenstein et al.[53] who report on electrically induced sharp magnetization switching in La$_{0.67}$Sr$_{0.33}$MnO$_3$/BaTiO$_3$(001) films. That approach differs in the respect of choosing a ferroelastic substrate with sharp and hysteretic strain changes. The magnitude of strain-modulated $M$ in La$_{0.7}$Sr$_{0.3}$MnO$_3$ appears comparable in both experiments even though the samples of Ref. 53 are in a less-defined, inhomogeneous strain state.

ACKNOWLEDGMENT

This work was supported by Deutsche Forschungsgemeinschaft, FOR 520. We are thankful to L. Eng , D. Hesse, M. Richter, M. Kuzmin and D. Lupascu for valuable discussions and to U. Keitel for interferometry measurements.

**Figure Captions**

FIG.1. In-plane piezoelectric strain vs. applied electric field $E \| [001]$ recorded along a [100] edge of a 0.72PMN-0.28PT(001) substrate.

FIG.2. Temperature dependence of the remanent magnetization $M \| [100]$ recorded at the indicated electric field $E \| [001]$ applied to the substrate. (a, b) Results of a La$_{0.7}$Sr$_{0.3}$MnO$_3$/PMN-PT(001) film, with a straight line fitted to the linear range of $M^2(T)$. (c, d) Respective results for a La$_{0.7}$Ca$_{0.3}$MnO$_3$/PMN-PT(001) film.

FIG.3. a) Magnetization $M \| [100]$ vs. electric field $E \| [001]$ applied to the substrate for a La$_{0.7}$Sr$_{0.3}$MnO$_3$/PMN-PT(001) film. b) Respective result for the remanent magnetization at unidirectional electric field for a La$_{0.7}$Ca$_{0.3}$MnO$_3$/PMN-PT(001) film. c) Magnetoelectric coupling coefficient $\alpha = \mu_0 \, dM/dE$ derived from the data of Fig. 3a.

FIG.4. Magnetic field ($H$) dependence of the magnetization $M$ of a La$_{0.7}$Sr$_{0.3}$MnO$_3$/PMN-PT(001) film recorded at the indicated electric field $E \|$ 001 applied to the substrate. $M_S$ denotes the saturation value at $\mu_0 H = $ 5 T and $E = $ 0. a) Measurement along an easy axis at 10 K. b) Measurement along a hard axis at 10 K. c) Measurement at 270 K.

FIG.5. Magnetization $M \| [100]$ vs. total in-plane strain of a La$_{0.7}$Sr$_{0.3}$MnO$_3$/PMN-PT(001) film.

FIG.6. Ferromagnetic Curie temperature ($T_C$) (a) vs. substrate electric field $E \| [001]$ and (b) vs. total in-plane strain for a La$_{0.7}$Sr$_{0.3}$MnO$_3$/PMN-PT(001) film. $T_C$ is derived from temperature-dependent magnetization data (cf. linear fits in Figs. 2b,d).

TAB.1: Studied films on PMN-PT(001) substrates, their out-of-plane lattice parameter $c$, perpendicular film strain $\varepsilon_{zz}$ in the as-grown state (derived using bulk parameters $a_{LSMO}$ = 3.876 Å and $a_{LCMO}$ = 3.864 Å), ferromagnetic Curie temperature $T_C$ in the as-grown state, and increase $\Delta T_C$ recorded upon application of a piezoelectric in-plane strain $\delta\varepsilon_{xx}$.

| Sample | Compound | Film thickness (nm) | $c$ (Å) | $\varepsilon_{zz}$ (%) | $T_C$ (K) | $\delta\varepsilon_{xx}$ (%) | $\Delta T_C$ ($\delta\varepsilon_{xx}$) (K) |
|---|---|---|---|---|---|---|---|
| #1 | La$_{0.7}$Sr$_{0.3}$MnO$_3$ | 50 | 3.840 | -0.93 | 339 | -0.11 | 7 |
| #2 | La$_{0.7}$Sr$_{0.3}$MnO$_3$ | 20 | 3.828 | -1.24 | 278 | -0.08 | 19 |
| #3 | La$_{0.7}$Sr$_{0.3}$MnO$_3$ | 30 | 3.850 | -0.67 | 340 | - | - |
| #4 | La$_{0.7}$Ca$_{0.3}$MnO$_3$ | 30 | 3.829 | -0.91 | 198 | -0.11 | 11 |

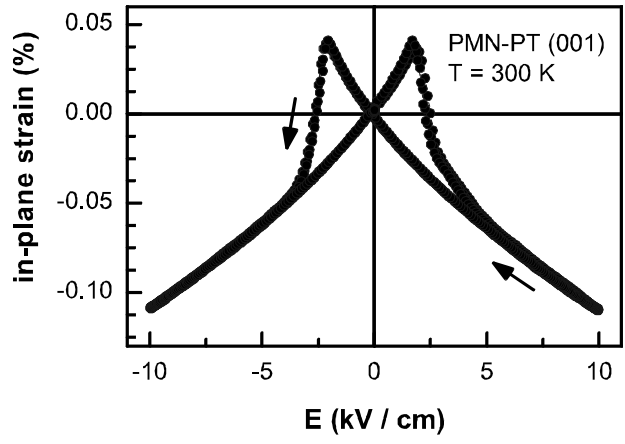

THIELE et al. FIG.1.

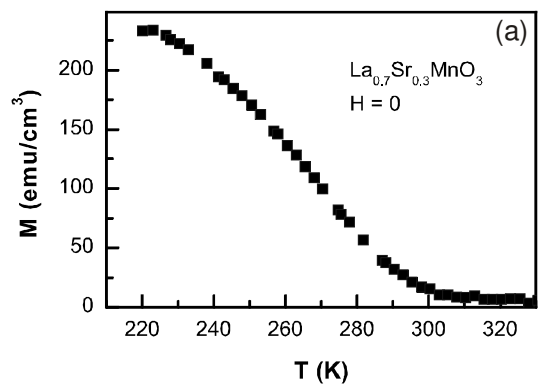
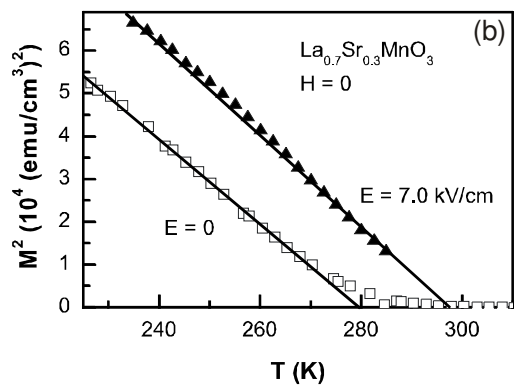
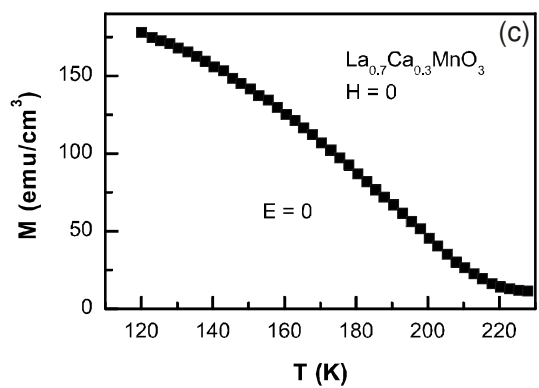
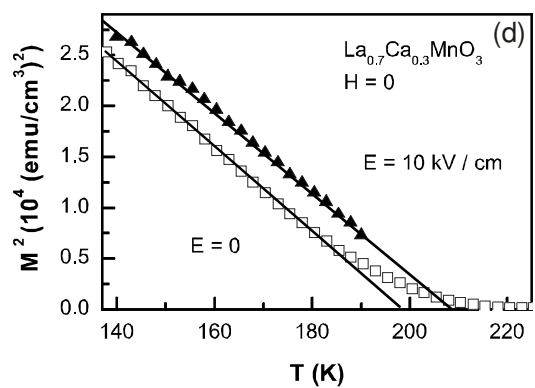

THIELE et al. FIG.2.

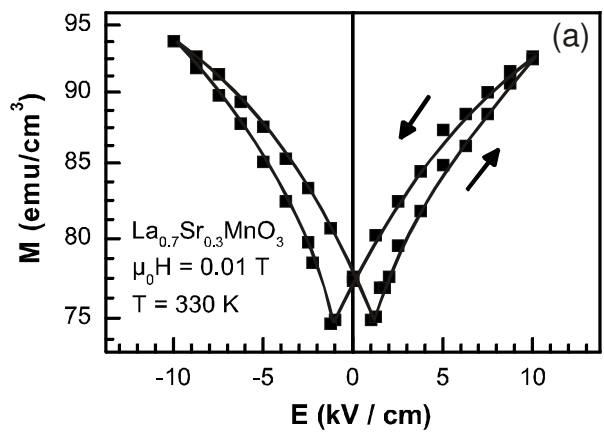
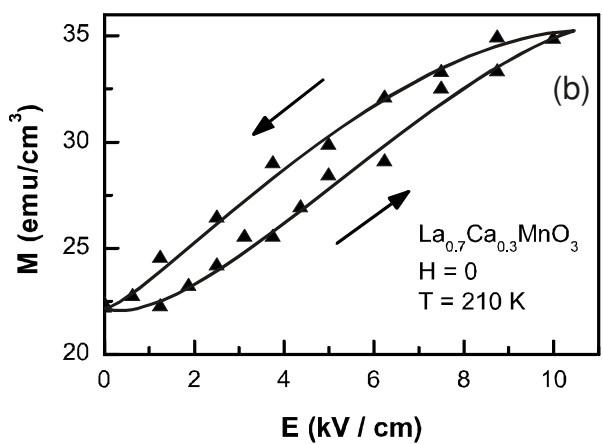
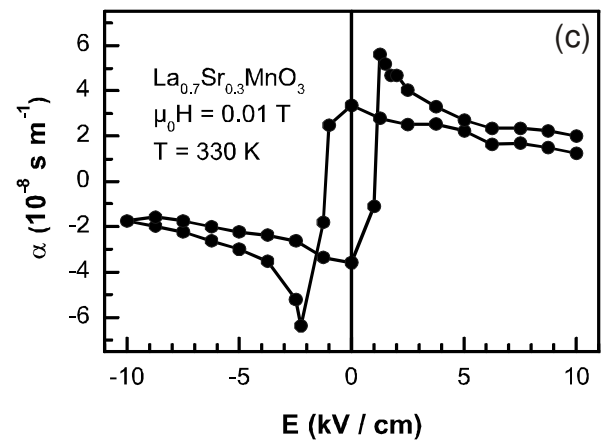

THIELE et al. FIG.3.

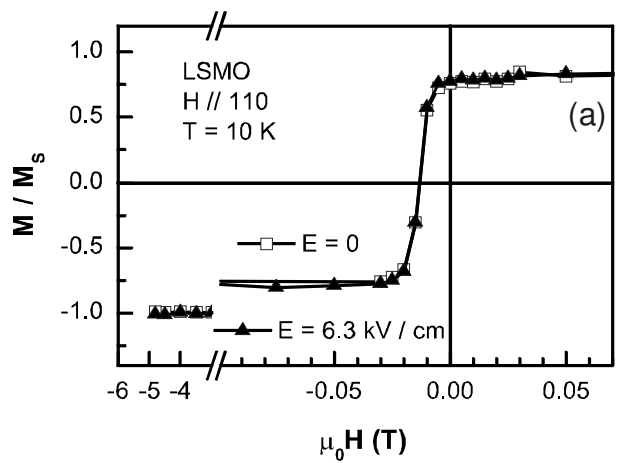
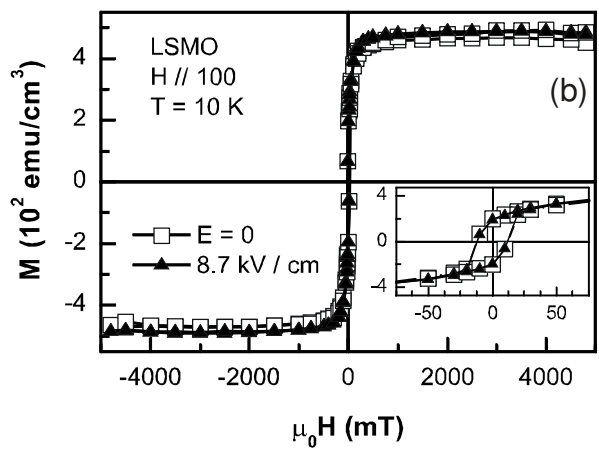
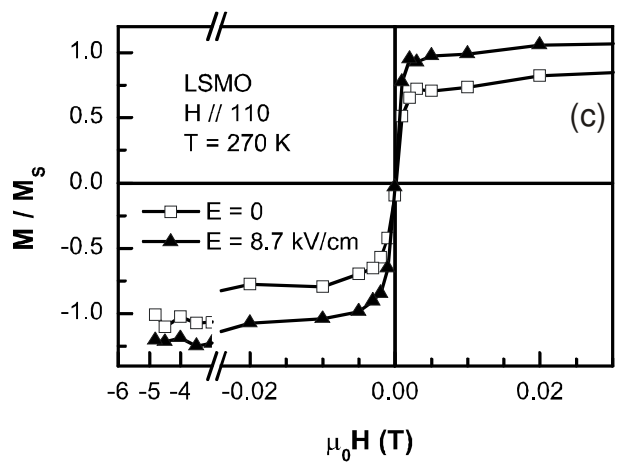

THIELE et al. FIG.4.

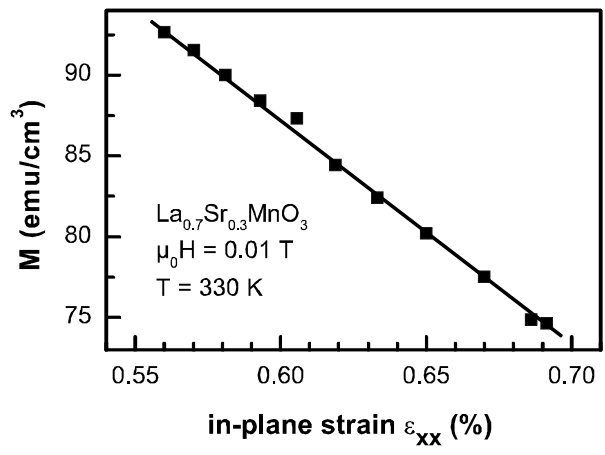

THIELE et al. FIG.5.

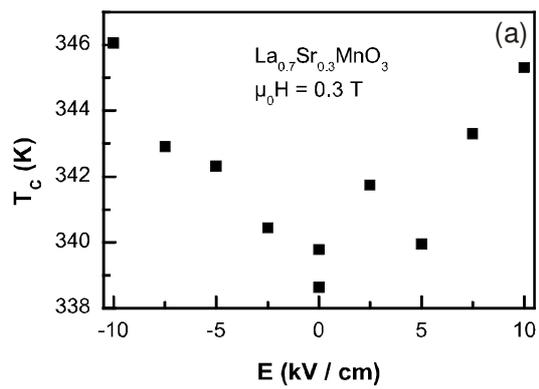 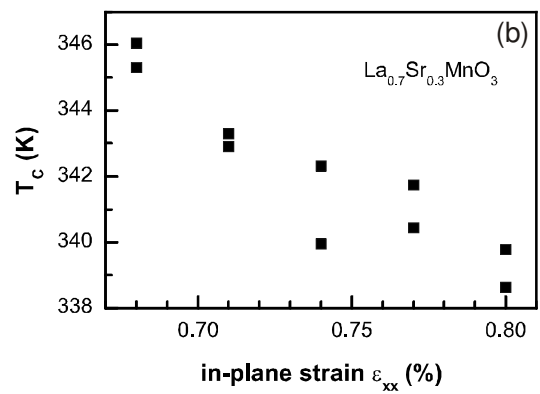

THIELE et al. FIG.6.